\documentclass[12pt]{iopart}
\usepackage{fullpage,graphicx,epsf}
\begin{document}
\title{A dynamically extending exclusion process}
\author{K. E. P. Sugden and  M. R. Evans}

\address{
SUPA, School of Physics, University of Edinburgh, \\
Mayfield Road, Edinburgh EH9 3JZ, UK.\\
}
\ead{k.e.p.sugden@sms.ed.ac.uk, m.evans@ed.ac.uk}

\begin{abstract}
An extension of the totally asymmetric exclusion process, which
incorporates a dynamically extending lattice is explored.  Although
originally inspired as a model for filamentous fungal growth, here the
dynamically extending exclusion process (DEEP) is studied in its own
right, as a nontrivial addition to the class of nonequilibrium
exclusion process models.  Here we discuss various mean-field
approximation schemes and elucidate the steady state behaviour of the
model and its associated phase diagram.  Of particular note is that
the dynamics of the extending lattice leads to a new region in the
phase diagram in which a shock discontinuity in the density travels
forward with a velocity that is lower than the velocity of the tip of
the lattice.  Thus in this region the shock recedes from both
boundaries.
\end{abstract}

\date{\today}
\maketitle


\section{Introduction}
\label{sec:intro}
When studying the physics of systems from the natural world, it is not
surprising that one finds many problems which lie outside the realm of
equilibrium physics.  Rather, we are more likely to find systems which
 reach some sort of nonequilibrium steady state, characterised by non
zero energy flow or other currents.  For studying such systems, a
vibrant collection of models has emerged.  Many of the models have
been studied extensively, not only for their applications but also for
their intrinsically interesting behaviour \cite{Mukamel}.

The asymmetric simple exclusion process (ASEP) is a paradigmatic model
for simple nonequilibrium driven diffusive systems \cite{BE07}.  It
comprises a one dimensional lattice along which hard-core particles
hop, with some bias in direction.  Despite its simplicity, this model
is capable of showing a range of interesting phenomena including
boundary induced phase transitions, shock formation and spontaneous
symmetry breaking.  Over the years the application of the ASEP has
been wide and varied, but it particularly lends itself to the
modelling of transport problems \cite{Chowdhury}, for example the
motion of molecular motors along microtubule filaments in biophysical
systems \cite{AMP,KLN,KL,FPF}.  Indeed the ASEP was originally
introduced as a lattice model of ribosome motion along mRNA
\cite{MGP}.

Recently, a particular case of the ASEP; the totally asymmetric
exclusion process (TASEP), was generalised to incorporate a
dynamically extending lattice \cite{SEPR,ES}.  The new variation
allows a particle reaching the end of a lattice to extend it by
converting into a single new site.  In this way, a connection is made
between the microscopic constituents of the model and the dynamics of
the system size.  We also note that very recently an asymmetric
exclusion process with Langmuir kinetics and a dynamic boundary has
also been studied \cite{NFC}.

The model of \cite{SEPR,ES} was motivated by a problem in mycology:
the process of filamentous fungal growth, where continuous growth of
the filament tip is maintained by a supply of mass transported from
behind the tip to the site of growth. In \cite{SEPR} it was argued
that one could model the transport of vesicles by molecular motors
along the network of microtubule fragments by an ensemble of
continuous effective microtubules acting as dynamically extending
asymmetric exclusion processes.  Preliminary mean-field and simulation
results showed that various growth regimes occurred, in particular
regimes exhibiting a high density of motors near the tip.

In this work, the extending TASEP of \cite{SEPR} is discussed in its
own right, as a nontrivial addition to the existing class of exclusion
process models.  We discuss how steady states may arise in this
growing system, and extend the well established TASEP phase diagram
into a 3-dimensional parameter space.  As with the TASEP, we may
derive the phase diagram through a phenomenological approach which
considers the dynamics of a \emph{shock} that travels between the
boundaries of the lattice.  Through this approach, we arrive at a new
subregion in the phase diagram in which a shock is moving away from
both boundaries and thus persists in the system.  We develop a series
of mean-field approximations for our analysis.  The simplest scheme,
which for the ASEP actually predicts the correct phase diagram
\cite{DDM}, only predicts the phases qualitatively. However a
refined mean-field approximation predicts phase boundaries which are
rather close to those we determine by Monte Carlo simulations.
A heuristic argument which considers the symmetry between particles
entering one end of the lattice and `holes' entering the opposite end
provides a further prediction for the phase diagram which is very
close indeed to that predicted by the refined mean-field theory.  We
find that this prediction is in fact the best fit to the simulation
data.

The paper is organised as follows: In section \ref{sec:tasep}, we
review the TASEP and introduce the dynamically extending model.  In
section \ref{sec:deep} the model is formally defined and in section
\ref{sec:mf} a simple mean-field approximation is used to analytically
approximate the steady states.  In section \ref{sec:impmf} the
mean-field approximation is extended to include correlations between
sites 1 and 2, resulting in a refined prediction of the model's phase
boundaries.
In section \ref{sec:heur}, we discuss the heuristic argument
and the resulting phase diagram which appears to be in closest
agreement with simulation.  The Monte Carlo simulation results are
discussed in section \ref{sec:pd}, where phase diagrams for the model
are also presented.  Finally, we conclude in section \ref{sec:conc}.


\begin{figure}
\centering
\includegraphics[scale=0.7]{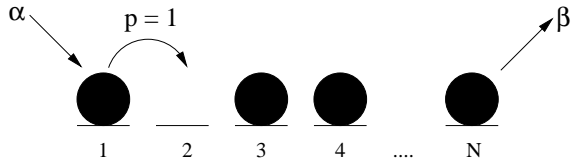}
\caption{\label{tasep} Schematic of the open boundary TASEP with input
rate $\alpha$, hop rate $p=1$ and extraction rate $\beta$.}
\end{figure}

\begin{figure}
\centering
\includegraphics[scale=0.5]{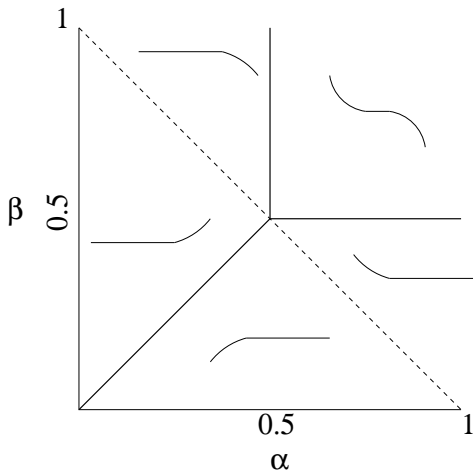}
\caption{\label{tasepPD} The phase diagram for the TASEP with schematic
  density profiles.  Three distinct phases exist.  For $\alpha$
  or $\beta<0.5$, two phases are possible; a high density phase for
  $\alpha>\beta$ and a low density phase for $\beta>\alpha$.  When
  $\alpha$ and $\beta>0.5$, the system is in a phase of maximum current,
  no longer limited by the input or output rate. }
\end{figure}

\begin{figure}
\centering
\includegraphics[scale=0.7]{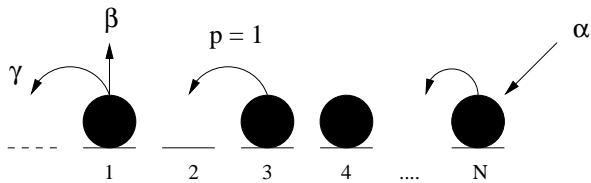}
\caption{\label{model} Schematic of the model with input rate
$\alpha$, hop rate $p=1$, desorption rate $\beta$ and growth rate
$\gamma$.}
\end{figure}


\section{A TASEP based model}
\label{sec:tasep}
We begin by reviewing the well-known TASEP, upon which the dynamically
extending exclusion process we study  is based.  It comprises particles
that hop along a one-dimensional lattice in a single direction.  No
more than one particle may occupy each lattice site at any given time.
Particles are injected at one end of the lattice and removed from the
other end with rates $\alpha$ and $\beta$, which are the only control
parameters for the model.  In the bulk of the lattice, the hops occur
with rate 1.  Figure~\ref{tasep} illustrates the TASEP schematically.

The behaviour of this system is categorised in terms of its
\emph{steady state phases}.  A steady state is reached when the
density of each site, that is the time-averaged occupancy, no longer
changes with time.  Equivalently, when the system is in a steady
state, the current of particles along the lattice is the same
everywhere.  The phases are defined by the system's macroscopic
properties in the large system limit i.e. the particle density
\emph{profile} across the lattice and the particle current between any
two neighbouring lattice sites.  Phase transitions are induced by
varying the boundary conditions, controlled by parameters $\alpha$ and
$\beta$.  The exact phase behaviour of the TASEP is well understood
from the exact solution \cite{DEHP,SD} as well as mean-field
\cite{DDM} and other approaches \cite{BE07}.  Three distinct steady
state phases exist.  For low input, $\alpha<0.5$, and $\alpha<\beta$,
the system is in a low density phase, where the bulk of the lattice is
at a low density equal to $\alpha$ and the current is equal to
$\alpha(1-\alpha)$.  When at $\alpha=\beta$, there is a discontinuous
phase transition and the system enters a high density phase, now
limited by a low output rate $\beta<0.5$.  The bulk of the lattice in
this phase is at a high density equal to $1-\beta$ and the current is
equal to $\beta(1-\beta)$.  When $\alpha$ and $\beta$ are both $>0.5$,
the system enters a phase of maximal current.  The system is no longer
controlled by the input and output rates, instead the bulk density is
$1/2$ and the current is $1/4$.  Furthermore, the density profile decays
algebraically from the boundaries towards the bulk value $1/2$.  The
results are summarised in the phase diagram of Figure~\ref{tasepPD}.

The model discussed in the following work has an additional third
parameter, $\gamma$, which is the rate at which a particle reaching
the ultimate lattice site may convert into a new lattice site, thus
extending the length of the lattice by 1.  We shall explore the
resultant phase structure in the three dimensional parameter space of
$\alpha$, $\beta$ and $\gamma$.

\section{The DEEP}
\label{sec:deep}
The dynamically extending exclusion process (DEEP) is defined by the
rates at which the following processes occur: particles in the bulk
hop to the left with rate 1; particles enter the lattice from the
right with rate $\alpha$; at site 1 two processes may occur, particles
detach from site $1$ with rate $\beta$ and particles may transform
into a new lattice site with rate $\gamma$.  These processes are
illustrated schematically in Figure~\ref{model}.  Thus $\gamma$ is the
parameter controlling the lattice growth, $\beta$ allows particles to
leave the end of the lattice without extending it and ratio
$\gamma/\beta$ controls the efficiency with which the lattice extends.

Indicating the presence of a particle by 1 and an empty lattice site
by 0 the stochastic dynamics is
\begin{eqnarray}\label{definition1}
& 01 \to 10&\quad\mbox{with rate}\quad 1\;,\\ \mbox{at site 1}\quad &1
\to 0 &\quad\mbox{with rate}\quad \beta\\ \mbox{at site 1}\quad & 1
\to 00 & \quad\mbox{with rate}\quad \gamma\;,\label{growth}\\
\mbox{at the rightmost site}\quad &0 \to 1
&\quad\mbox{with rate}\quad \alpha \;.
\label{alpha}
\end{eqnarray}
Note the direction of particle hopping is to the {\em left} (as
opposed to most studies of the TASEP where the direction is to the
right) for reasons which will soon become apparent.  This choice is of
course immaterial.

Also note that under the simultaneous reversal of the direction of
particle hopping and interchange of $\alpha$ and $\beta$, the TASEP
enjoys an exact particle-hole symmetry at the microscopic level, while
the DEEP does not.  This is due to `growth' rule \ref{growth}, which
has no counterpart at the right boundary.

\subsection{Exact steady state  equations}

In the DEEP model the system size is not conserved and continually
increases, therefore care must be taken to define what is meant by a
steady state.  For example, the density at site $i$ will generally
only become stationary with respect to a particular frame of
reference.  Two natural frames to consider are the frame moving with
the tip, in which case the tip is always at site 1 and the site label
$i$ measures the distance from tip, and the stationary frame where the
right boundary is always at site 1 and the tip is at site $N$ where
$N$ increases as the lattice extends. A phase of the system is
specified by stationary values of the densities and the frame of
reference. Moreover we shall encounter one phase, the maximal current
phase, that does not satisfy this criterion and instead is only
quasi-stationary.

To begin with, we choose to work in the reference frame of the growing
lattice tip, so that the leftmost site is always labelled site 1.  The
input end of the lattice is site $N(t)$ and process (\ref{alpha}) may
be formulated as there being a reservoir of particles of density
$\alpha$ at site $N(t)+1$.  As time $t\to \infty$ the system size
$N(t) \to \infty$, therefore we may write a boundary condition at the
input end of the lattice as
\begin{equation}\label{BC}
\alpha = \lim_{N \to \infty} \langle\tau_N\rangle\;.
\end{equation}

We now construct the equations for the steady state dynamics.  Note
that when a growth event occurs, via process (\ref{growth}), all the
previous lattice sites are relabelled $i\ \rightarrow\ i+1$.  The
exact equations for correlations functions in the steady state are:
\begin{eqnarray}
\label{tau1}
\frac{d\langle\tau_1\rangle}{dt}=0=\langle[1-\tau_1]\tau_{2}\rangle-(\gamma+\beta)\langle\tau_1\rangle\;,\\
\label{tau2}
\frac{d\langle\tau_2\rangle}{dt}=0=\langle[1-\tau_2]\tau_{3}\rangle-\langle[1-\tau_{1}]\tau_2\rangle-\gamma\langle\tau_1\tau_2\rangle\;,\\
\label{taui}
\frac{d\langle\tau_i\rangle}{dt}=0=\langle[1-\tau_i]\tau_{i+1}\rangle-\langle[1-\tau_{i-1}]\tau_i\rangle+\gamma\langle\tau_1[\tau_{i-1}-\tau_i]\rangle\quad
i\geq 3\;.
\end{eqnarray}
Where $\tau_i$ is the occupancy at site i and can take on a value 1 or
0. The angle brackets denote an average in the steady state.  Thus
$\langle \tau_i \rangle$ is the average occupancy in the steady state
or particle \emph{density} at site i.  These equations may be
understood in terms of a particle \emph{current} along the lattice.
The positive terms represent the current of particles entering the
site from the right and the negative terms represent the current
leaving to the left.  The dynamics at sites 1 and 2 differ from those
of the rest of the system due to the transition from particle
to lattice site.  In the steady state, the current `in'
balances the current `out' everywhere.  We thus obtain from
(\ref{tau1}-\ref{taui}) the following exact expressions for the
conserved steady state particle current $J$:
\begin{eqnarray}
\label{v1}
J=(\gamma+\beta)\langle\tau_1\rangle\;,\\
\label{v2}
J=\langle\tau_2[1-\tau_{1}]\rangle\;,\\
\label{vi}
J=\langle\tau_{i+1}[1-\tau_{i}]\rangle-\gamma\langle\tau_1\tau_{i}\rangle\quad
i\geq 2\;.
\end{eqnarray}
The rate of lattice extension, or tip velocity, is given by:
\begin{equation}
\label{v}
v=\gamma\langle\tau_1\rangle\;,
\end{equation}
which leads us to a relation between the current and velocity:
\begin{equation}\label{jv}
J=v(1+\frac{\beta}{\gamma}).
\end{equation}
Our approach to solve the system of equations (\ref{v1}--\ref{vi}) is
to take a mean field approximation in order to analyse the steady
state equations and determine the phases and phase boundaries of the
system.

\section{Simple mean-field theory}
\label{sec:mf}
In the simplest mean field approximation, we replace
\begin{equation}
\langle \tau_i \tau_j \rangle = \rho_i \rho_j \qquad \forall i,j\; (i\neq j)
\end{equation}
where $\rho_i = \langle \tau_i \rangle$ is the average steady state density at site $i$, and we ignore
correlations between the density at different sites \cite{DDM}.
Equation (\ref{vi}) then gives the recurrence relation:
\begin{equation}\label{RR}
\rho_{i+1}=\frac{J+v\rho_{i}}{1-\rho_{i}}\quad i\geq 2\;.
\label{recur}
\end{equation}
We find $\rho_1$, $\rho_2$ from (\ref{v1}), (\ref{v2}) and (\ref{jv}):
\begin{equation}\label{rho1}
\rho_1=\frac{J}{\gamma+\beta}
\end{equation}
\begin{equation}
\rho_2=\frac{J}{1-\rho_1}=\frac{J(\gamma+\beta)}{\gamma+\beta-J}\;.
\label{rho2}
\end{equation}
The recurrence (\ref{recur}) together with the initial condition
(\ref{rho2}) determine all $\rho_i$ in terms of $J$ (or $v$).
It remains to fix $J$ self-consistently to determine the various phases.
\vspace*{1ex}

\begin{figure}
\centering
\includegraphics[scale=0.45]{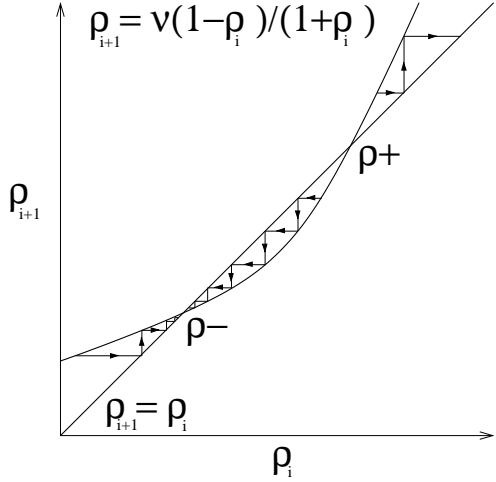}
\caption{\label{iterations}A graphical representation of the solutions
  to the recurrence relation (\ref{RR}).  Iterations beginning below
  the unstable upper fixed point, $\rho_+$, converge to the stable
  lower fixed point, $\rho_-$.}
\end{figure}

\noindent {\em Low density Phase:}
The recurrence (\ref{recur}) has two fixed points
\begin{equation}\label{fp}
\rho_{\pm}=\frac{1-v\pm\sqrt{(1-v)^2-4J}}{2}\;.
\end{equation}
The lower of the fixed points is stable, so for $\rho_2$ within the
basin of attraction, the density iterates to this value and the
density profile converges rapidly to a constant density in the bulk of
the lattice.  See Figure~\ref{iterations} for a graphical
representation of the iterative solutions to (\ref{RR}).  To meet
boundary condition (\ref{BC}), we define the lower fixed point:
\begin{equation}
\label{lfp}
\rho_-=\alpha\;.
\end{equation}
It follows from (\ref{fp}) that the upper fixed point is then:
\begin{equation}
\label{ufp}
\rho_+=1-v-\alpha\;.
\end{equation}

The boundary condition (\ref{BC})
fixes $v$ using (\ref{jv}) and (\ref{RR}):
\begin{equation}
v_{ld}=\frac{\alpha(1-\alpha)}{1+\alpha+\frac{\beta}{\gamma}}\;,
\label{vld}
\end{equation}
and the corresponding current is
$J_{ld} = (1 + \beta/\gamma) v_{ld}$.

The condition for this phase is that $\rho_2<\rho_+$ so that
$\rho_i\rightarrow\rho_-$ as $i\to \infty$.  Using (\ref{ufp}) and (\ref{vld}) this
condition reduces to
\begin{equation}
\gamma>\frac{\alpha-\beta}{1+\alpha}\;.
\label{ldcon}
\end{equation}
This is the phase boundary for the low density phase.

The density in the bulk of the low density phase is limited by a low
input rate $\alpha$.  At the tip there is a decay to the boundary
condition (\ref{lfp}) determined by rates $\beta$ and $\gamma$.  The low density
region is split into two subregions distinguished by whether the tip
density is above or below $\alpha$.  For $\rho_2<\alpha$, the tip
density is less than $\alpha$ and for $\alpha<\rho_2<1-v-\alpha$, the
tip density is greater than $\alpha$.

\vspace*{1ex}
\begin{figure}
\centering
\includegraphics[scale=0.5]{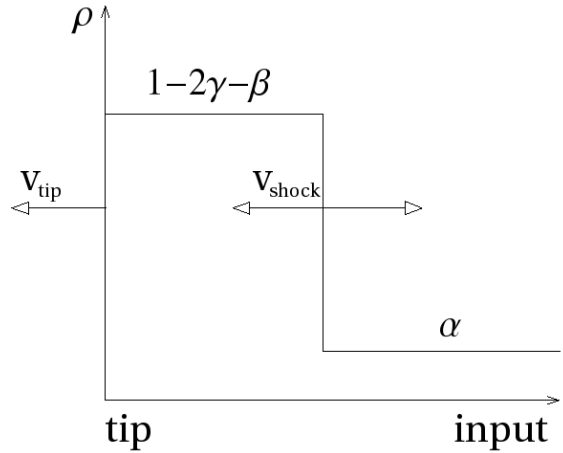}
\caption{\label{shock}A shock is set up between regions in the lattice with a
  high density (next to the tip) and low density (far from the tip).
  The shock travels with a velocity fixed through mass conservation,
  which may be different to the velocity of the lattice tip.}
\end{figure}
\vspace*{1ex}

\noindent {\em High density phase:} Another possible solution to
(\ref{recur}) and (\ref{BC}) is that $\rho_2 = \rho_+$.  In this case
the recurrence (\ref{recur}) yields $\rho_i = \rho_+$ for $i \geq 2$.
The phenomenological interpretation is that when $\gamma$, the rate of
release of particles at the growing end, is no longer large enough to
control the input rate; the particle density near the tip reaches a
maximum value $\rho_+$ that extends back through the lattice.  The
high density in the bulk $\rho_+$ is then connected to the density at
the input boundary $\alpha$, with a discontinuous \emph{shock}. Mass
conservation implies a speed for the shock, $v_s$, given by:
\begin{equation}\label{vshock}
v_s =  1- \rho_+ -\alpha\;.
\end{equation}
The shock solution is illustrated schematically in Figure~\ref{shock}.

Defining $\rho_2=\rho_+$, we obtain from (\ref{jv}), (\ref{rho2}) and
(\ref{fp}), the tip velocity in this phase:
\begin{equation}
\label{vhd}v_{hd}=\frac{\gamma(1-2\gamma-\beta)}{1-\gamma}\;,
\end{equation}
and from this and (\ref{ufp}), we have an expression for the density
at site $2$ and throughout the rest of the high density region:
\begin{equation}
\label{p+}
\rho_+=1-2\gamma-\beta\;.
\end{equation}
From (\ref{ldcon}), one can see that this phase is entered when:
\begin{equation}
\gamma=(\alpha-\beta)/(1+\alpha)\;.  
\end{equation}
Due to the high density structure in the tip region we refer to the
phase as the \emph{`jammed'} or \emph{high density phase}.  Note
however that only on the phase boundary where $v_s=v$, is the shock
and therefore the density profile stationary in the reference frame of
the tip.  In the rest of this phase $v_s < v$, and the shock moves
away from the tip, resulting in an expanding high density region.
Provided $v_s> 0$, the shock will also be moving away from the right
boundary and the expanding high density region will never occupy the
whole system.  Using conditions (\ref{fp}) and (\ref{vshock}), the
condition $v_s>0$ becomes:
\begin{equation}
\label{shockreg}
\frac{\alpha-\beta}{1+\alpha}>\gamma > \frac{\alpha-\beta}{2}\;.
\end{equation}
In order to fulfil our criterion for a steady state (stationary densities
in our frame of reference as $t\to \infty$) we should
be in the reference frame of the shock. Then  the density in front of the shock
is given by $1-2\gamma -\beta$ and the density behind is given by $\alpha$.

On the other hand if $v_s<0$, then the shock moves backwards through
the system until it reaches the right boundary.  In this region a high
density is maintained throughout the length of the lattice except for
a boundary region next to the right boundary. The system may be
considered to be in a high density steady state with the reference
frame being the stationary frame where the right hand boundary is
fixed.  The condition $v_s<0$ becomes
\begin{equation}
\label{hdreg}
\gamma<\frac{\alpha-\beta}{2}\;.
\end{equation}
\vspace*{1ex}

\noindent {\em Maximal current phase}:
Both the high and low density phases are bounded by the condition that the
roots $\rho_\pm$ of (\ref{RR}) are real.  Clearly, the roots (\ref{fp})
are real provided
$(1-v)^2>4J$, which implies  using (\ref{jv}), that 
$v< v_{max}$, where the 
maximum velocity satisfies
\begin{eqnarray}
\label{maxV}
v_{max}&=&\frac{2\beta+3\gamma-2\sqrt{\beta^2+3\gamma\beta
+2\gamma^2}}{\gamma}\;.
\end{eqnarray}  
and the  corresponding  maximum current
is $J_{max} = (1+ \beta/\gamma)v_{max}$.

When $J=J_{max}$,  we have a single  root of (\ref{RR}) 
and from  (\ref{fp}) we deduce that the bulk density is
\begin{equation}
\rho_{max} = \frac{1-v_{max}}{2}\;.
\end{equation}
We refer to this phase as the maximal current phase; the flow is no longer
controlled by the boundary condition $\alpha$ rather it has saturated at a 
maximal flow rate, $J_{max}$.   Note that $\rho_{max} <1/2$ and
$J_{max} <1/4$.

On a finite system of size $N$, in order to fix the boundary
conditions, one takes $J= J_{max} + O(1/N^2)$ \cite{DDM,BE07}.  The
iterative solution of (\ref{recur}) then yields a density profile
decaying smoothly between both boundaries with an algebraic rather
than an exponential decay of the density from the boundaries.  Note
that there is thus no natural reference frame in which we may define a
steady state.  Rather, the density at each site evolves as the system
grows and we characterize this phase as quasi-stationary.  We find the
phase boundaries by considering the transitions to the maximal current
phase from the low density phase, and from the high density phase:
$J_{ld}=J_{max}$ when
\begin{equation}\label{maxG}
\gamma=\frac{\beta-2\alpha\beta}{\alpha^2-1+2\alpha}\;,
\end{equation}
and $J_{hd}=J_{max}$ when
\begin{equation}\label{maxB2}
\beta=\frac{4\gamma-2\gamma^2-1}{\gamma-2}\;.
\end{equation}
\vspace*{1ex}

\begin{figure}
\centering
\includegraphics[scale=0.5]{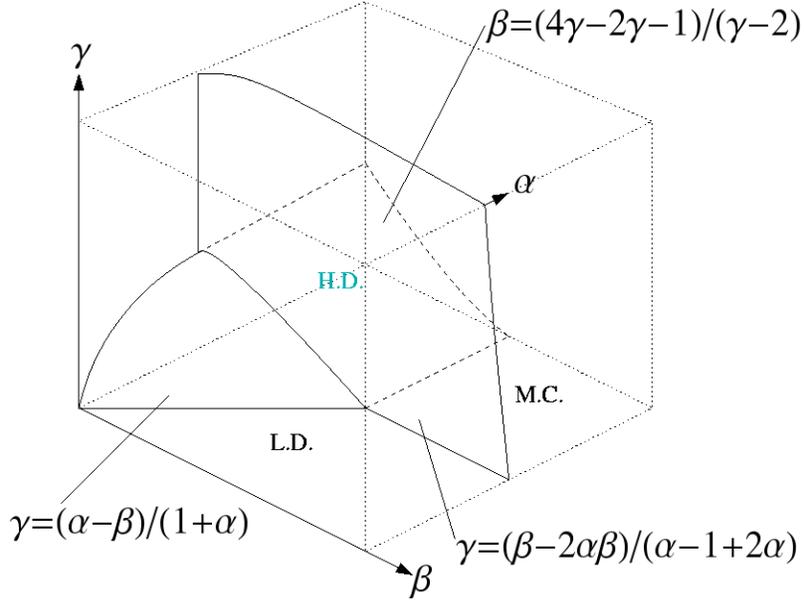}
\caption{\label{3DSimple}A schematic representation of the phase
  diagram in 3-D $\alpha,\ \beta,\ \gamma$ parameter space for the
  simple mean-field approximation.  The three phases are shown
  (without subregions).  The $\gamma=0$ plane corresponds to the well
  known TASEP phase diagram.}
\end{figure}

\noindent {\em 3D Phase diagram}: In order to visualise the phase
structure of the DEEP, we may construct a 3-dimensional phase diagram
from the results of the simple mean-field theory.
Figure~\ref{3DSimple} is a schematic representation of the three phase
regions, joined by the phase boundary surfaces described above.  On
the diagram, one can see that in the limit of no growth ($\gamma=0$),
the phase boundaries agree with those of the TASEP.

\subsection{The case $\beta=0$}
We may greatly simplify the equations of this model, by considering
the case $\beta=0$.  This is the limit of no extraction from the tip
and every particle reaching site 1 now contributes to the lattice
growth.  The current and velocity are therefore equal ($J=v$).  We now
have in the steady state:
\begin{eqnarray}
\rho_1 = v/\gamma \qquad \rho_2 = \gamma v/(\gamma- v)\\[1ex]
\rho_{i+1} = \frac{v(1+\rho_i)}{1-\rho_i}\quad i\geq 2\;. 
\label{recurb0}
\end{eqnarray}

The condition for the low density phase (\ref{ldcon})
reduces to $\gamma>\alpha/(1+\alpha)$ and  the tip velocity is
given by
\begin{equation}
v=J=\frac{\alpha(1-\alpha)}{(1+\alpha)}\;.
\label{vldb0}
\end{equation}
Then, the  condition that $\rho_\pm$ be real becomes
\begin{equation}
J=v<3-2\sqrt{2}\;,
\label{jmaxb0}
\end{equation}
which, with (\ref{vldb0}), implies
\begin{equation}
\alpha<\sqrt{2}-1\;.
\end{equation}

In the high density phase, where  $\gamma < \alpha/(1+\alpha)$,  the tip
velocity (\ref{vhd}) is
\begin{equation}
v=J=\frac{\gamma(1-2\gamma)}{1-\gamma}\;.
\label{vhdb0}
\end{equation}
For  $\alpha/2<\gamma<\alpha/(1+\alpha)$ a dynamic shock exists in
the system. 

In the high density phase condition  (\ref{jmaxb0}) 
together with (\ref{vhdb0}) imply
\begin{equation}
\gamma<1-\frac{1}{\sqrt{2}}\;.
\end{equation}
Therefore for the region bounded by
$\alpha>\sqrt{2}-1$, $\gamma>1-1/\sqrt{2}$, the system is in the
maximal current phase, where the bulk density is $\rho_{max}=\sqrt{2}-1$ and the current is $J_{max}=3-2\sqrt{2}$.

Monte-Carlo simulations of this system reveal that each of the
predicted phases are indeed present.  However, although the
qualitative results of this simple mean-field theory appear to be
correct, the predicted phase boundaries do not all agree with
simulation (as we shall discuss in Section \ref{sec:pd}).  Thus in
order to more accurately predict the phase boundaries, we develop a
refined mean-field theory.

\section{Refined mean-field theory}
\label{sec:impmf}
In the approximation of the previous section, we ignored correlations
between all pairs of sites in this system.  
In this section we introduce a refined mean field approximation
which retains information about some correlations.

We might expect  that sites
1 and 2 are in fact strongly correlated because when site 1 is
vacated, so too is site 2.  We can hope to  improve our approximation
by keeping the correlation $\langle\tau_1\tau_2\rangle$ intact in
(\ref{tau1}-\ref{v1}) and beginning the iteration of (\ref{RR}) at
site 3 instead of 2. Thus we approximate, for example,
\begin{equation}
\langle \tau_1 \tau_2  \tau_i \rangle
= \langle \tau_1 \tau_2  \rangle \rho_i\quad i\ge 3.
\end{equation}

To begin the recurrence  we must find a new expression for $\rho_3$
without factorising  $\langle \tau_1 \tau_2  \rangle$, for example.
Within the approximation, we have from (\ref{vi}),
\[\rho_3=\frac{J+\gamma\langle\tau_1\tau_2\rangle}{1-\langle\tau_2\rangle}\]
and from (\ref{v2})
\[\langle\tau_2\rangle=J+\langle\tau_1\tau_2\rangle\;.\]
Combining these two relations gives
\begin{equation}
\label{correlation}
\langle\tau_1\tau_2\rangle=\frac{\rho_3[1-J]-J}{\gamma+\rho_3}\;.
\end{equation}
 
Now we consider the steady state rates into and out of the
configuration in which site 1 is vacant and site 2 is occupied:
\begin{equation}
\label{tau12}
\frac{d\langle\tau_2[1-\tau_1]\rangle}{dt}=0=\langle\tau_{3}[1-\tau_2][1-\tau_1]\rangle-\langle\tau_2[1-\tau_1]\rangle+\beta\langle\tau_1\tau_2\rangle\;.
\end{equation}
We can identify the second term as the exact current from
(\ref{v2}).  
Thus we obtain another exact expression for the steady state current:
\begin{equation}
\label{v12}
J=\langle\tau_{3}[1-\tau_2][1-\tau_1]\rangle+\beta\langle\tau_1\tau_2\rangle\;.
\end{equation}
Factorising beyond site 2,
we may simplify and rearrange to give
\begin{equation}
\label{T3}
\rho_3=\frac{J-\beta\langle\tau_1\tau_2\rangle}{1-J-\frac{J}{\gamma+\beta}}\;,
\end{equation}
which along with (\ref{correlation}), gives us a quadratic equation for
$\rho_3$
\begin{equation}
\label{quadratic}
\rho_3^2[1-J-\frac{J}{\gamma+\beta}]+\rho_3[(\gamma+\beta)(1-J)-\frac{\gamma
     J}{\gamma+\beta}-J]-(\beta+\gamma)J=0\;.
\end{equation}
This may be solved straightforwardly and used to deduce the three dimensional phase surfaces as in section \ref{sec:mf}.

\subsection{The case $\beta=0$}
We consider again the simple case of $\beta=0$.  Then (\ref{T3}) reduces to
\begin{equation}
\rho_3=\frac{J}{1-J-\frac{J}{\gamma}}
\end{equation}
The condition for a low density phase, $\rho_3<1-v-\alpha$, is now
\begin{equation}
\alpha<\gamma
\end{equation}
and the velocity as before is $v=\alpha(1-\alpha)/(1+\alpha)$.  In the
high density phase, $\gamma<\alpha$ and the velocity is
$v=\gamma(1-\gamma)/(1+\gamma)$.  For
$\alpha/(2-\alpha)<\gamma<\alpha$ a dynamic shock exists between the
high density tip region $\rho=\rho_+=(1-\gamma)/(1+\gamma)$ and low
density region $\rho=\alpha$ far from the tip.  The maximum current
phase is attained when $\alpha>\sqrt{2}-1$ and $\gamma>\sqrt{2}-1$.

\subsection{Further correlations}
We can repeat this mean-field improvement scheme by taking into
account more correlations and correspondingly beginning the iteration
of (\ref{RR}) further from the tip.  One might expect that as further
correlations are taken into account, the predicted phase diagram will
converge toward the exact solution.  As we shall see however, this is
not always the case.  We do not present the details of the further
refined mean-field theory here as the equations become rather
cumbersome, but the results for the phase boundaries are shown in
Figure~\ref{2D-all} and discussed later, in Section \ref{sec:pd}.

\section{Symmetry and heuristic argument}\label{sec:heur} 
An important feature which allows us to understand the phase behaviour
of the TASEP (see Figure \ref{tasepPD}) is the presence of a symmetry
between particles hopping in one direction and `holes' hopping in the
opposite direction.  The high density particle phase can therefore be
understood as a low density hole phase and vice versa.

Due to the moving boundary condition and the transition
(\ref{growth}), there is no exact microscopic symmetry between
particles and holes in the DEEP dynamics.  However, it is true that
particles leave/holes enter at the tip whereas particles enter/holes
leave at the right end. It is plausible then that there exists some
symmetry between the high density phase (low density of holes) and the
low density phase (high density of holes).  Indeed, the phase
boundaries between the high, low and maximal current phases predicted
by the refined mean-field theory in the limit of $\beta=0$ display a
symmetry under the interchange $\alpha\leftrightarrow\gamma$.  But
while the the velocities in the two phases,
$\alpha(1-\alpha)/(1+\alpha)$ and $\gamma(1-\gamma)/(1+\gamma)$, are
also symmetric under the interchange $\alpha\leftrightarrow\gamma$ ,
the bulk densities in these phases, $\alpha$ and
$(1-\gamma)/(1+\gamma)$, are not symmetric.

In the case of $\beta\neq0$, the refined mean-field approximation no
longer displays a particle hole symmetry.  A heuristic argument
however, does lead to a 3D phase transition between the high and low
density phases which maintains a symmetry between input rate $\alpha$
and output rate $\beta+\gamma$.

If we ignore the density structure near the tip and assume that in the
high density phase there exists a single high density value associated
with the tip: $\rho_1=\rho_{bulk}$, in the same way that in the low
density phase, we have a single density $\alpha$ associated with the
input end of the lattice, then from (\ref{vi}-\ref{jv}) we have the
following:
\begin{equation}
v=\gamma\rho_1=\frac{\rho_1(1-\rho_1)}{1+\rho_1+\frac{\beta}{\gamma}}\;.
\end{equation}
This is solved to give the bulk density:
\begin{equation}
\rho_1=\rho_{bulk}=\frac{1-\gamma-\beta}{1+\gamma}\;,
\end{equation}
and associated high density velocity:
\begin{equation}
\label{v_heur}
v=\frac{\gamma(1-\gamma-\beta)}{1+\gamma}\;.
\end{equation}

As before we appeal to the concept of a shock discontinuity in the
system.  The low density phase is reached when the shock velocity
$(1-\rho_{+}-\alpha)$ is greater than the tip velocity (\ref{v_heur}),
which corresponds to a phase transition at:
\begin{equation}
\alpha=\beta+\gamma\;,
\end{equation}
suggesting an exact correspondence between the balancing of input and
output rates and the transition between high and low density phases.
This boundary is the same as that derived by the refined mean-field
theory in the limit $\beta=0$ and is in fact very close to that
derived in the case $\beta\neq0$ (see Figure \ref{fig:withBeta} in the
next Section).  The remaining phase boundaries are derived as in
section \ref{sec:mf}.

\section{Phase Diagrams and Simulation}
\label{sec:pd}
\begin{figure}
\vspace*{2ex}
\centering
\includegraphics[scale=0.4]{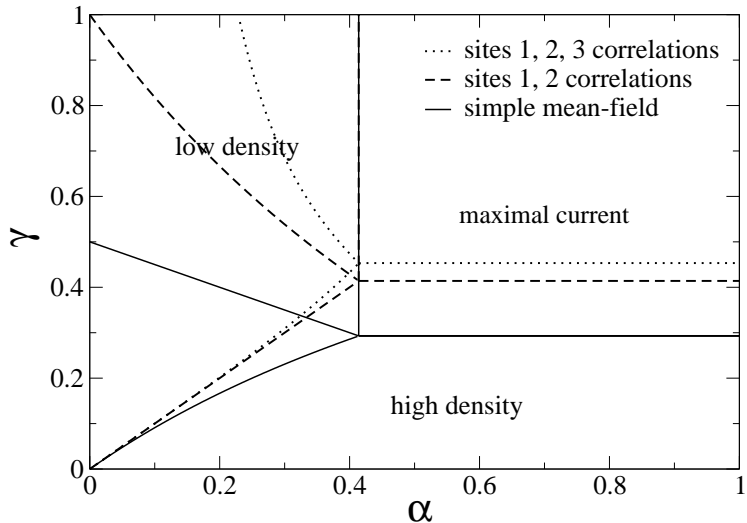}
\caption{\label{2D-all}Comparison of three mean-field approximations
in the case $\beta=0$.}
\end{figure}

As we have seen, the mean-field theories discussed above predict the
same phases but with differing phase boundaries.  The transition
between the low density and maximal current phases (\ref{maxG}) is the
same in each theory, however the transitions between other phases vary
significantly.  For comparison, the basic 2-D phase diagrams for the
simple, refined and further refined mean-field theories in the limit
of $\beta=0$ are plotted in Figure~\ref{2D-all}.  Note in particular,
a large difference in the boundary between subregions in the low
density phase where the tip density decays from above or below
$\alpha$.  The high to low density transition is also significantly
different in the simple mean-field theory.  In order to explore the
exact phase structure of the DEEP model, and identify which of the
theories best describes the exact behaviour of the model, we turn now
to a numerical discussion.  For simplicity, we work primarily in the
$\beta=0$ limit.  Note that in this limit the refined mean-field
theory of Section~\ref{sec:impmf} and the heuristic argument of
Section~\ref{sec:heur} are in agreement.

Monte-Carlo simulations of the DEEP were carried out across the
parameter space.  The simulated lattice comprises an array of binary
values representing $\tau_i$.  For each update, a site in the lattice
is selected randomly and a transition is attempted with a probability
defined by the rates (\ref{definition1}-\ref{alpha}).  One time-step
consists of $N$ such updates, so that on average each site is updated
once per time-step.  Thus as $N$ increases, so does the number of
updates per time step, in order to keep our time unit constant.  Note
that during one time-step the system may increase in length.  In
principle, this would affect our unit of time, however, such an
increase, typically of one lattice unit, will be insignificant in the
large $N$ limit.

Our aim is to allow the system to reach a steady state, and then we
may calculate quantities such as site densities, particle flux and
rate of change in lattice length.  A problem with this method is that
as the system grows, $N\rightarrow\infty$ and the computational cost
for each time-step increases.  We may however minimise this problem by
considering only a fixed length portion of the lattice in a particular
frame of reference.  For example, in the low density phase we expect
that the density tends exponentially to $\alpha$, so we may truncate
the lattice at site $M<N$ and simulate only the $M$ sites nearest the
tip.  This method will only work for the steady state phases which are
clearly defined in some reference frame.  For the maximal current
phase and shock region, we still consider the whole system.

\begin{figure}
\centering
\includegraphics[scale=0.4]{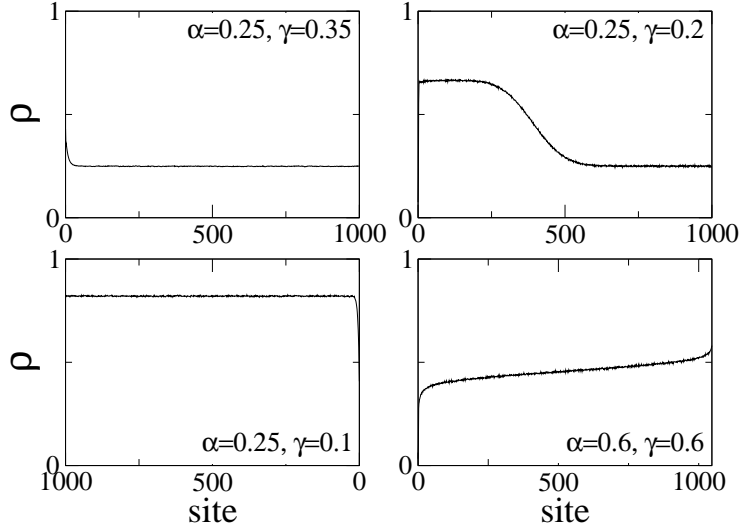}
\caption{\label{profs}Density profiles from the different phases of
  the dynamically extending exclusion process.  From top left: Low
  density, shock profile, high density and maximal current phase.
  Note that since the high density profile is calculated in the
  reference frame of the input, the sites are labelled from right to
  left, site 0 being where particles are injected.}
\end{figure}

An example of each type of density profile is plotted in
Figure~\ref{profs}.  In the low density phase, the steady state
densities are calculated by averaging the occupancy over many
time-steps in the reference frame of the tip.  On the other hand, in
the high density phase, the steady states are defined in the reference
frame of the input and the average occupancies are calculated from
this frame.  For profiles in the shock region and the maximal current
phase, the system is never in a truly stationary state, and to view
the density profile, we must find the average occupancies at a fixed
length and thus fixed time.  The densities here are calculated by
averaging over many arrays which have grown to a specified length.

\begin{figure}
\centering
\includegraphics[scale=0.4]{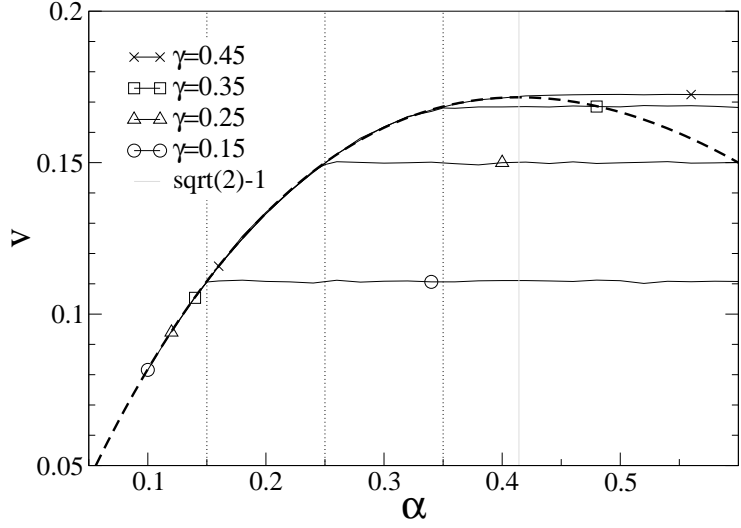}
\caption{\label{MontecarloV}The tip velocity plotted against $\alpha$
for $\gamma=0.15..0.45$ and $\beta=0$.  For $\gamma<\sqrt{2}-1$, a
clear phase transition occurs when $\alpha\approx\gamma$.  This
transition is from the \emph{low density} phase where
$v=\frac{\alpha(1-\alpha)}{1+\alpha}$ to the \emph{high density
phase}, where the velocity is $\alpha$ independent.  For clarity,
dotted lines are used to indicate $\alpha=\gamma$.  When
$\gamma=0.45>\sqrt{2}-1$, the maximum velocity occurs as predicted at
$\alpha=\sqrt{2}-1$.}
\end{figure}

One property which may be used to locate the exact positions of the phase
transitions is the lattice tip velocity, which is given by a simple
function of the parameters and changes form across the phase
boundaries (see equations (\ref{vld},\ref{vhd},\ref{maxV})).  We find
from the velocity of the simulated lattices, that the first refined
mean-field theory appears to predict the correct form for the tip
velocity and the correct phase transitions.  Figure~\ref{MontecarloV}
shows the velocity of a simulated lattice plotted for different values
of $\gamma$, against the refined mean-field expression for the
velocity.  The velocities closely follow the predicted curve
(\ref{vldb0}) and as expected, a phase transition takes place at
$\alpha=\gamma$, where the velocity becomes independent of $\alpha$.
The maximum velocity is reached when $\alpha=\sqrt{2}-1$ and
$\gamma\ge\sqrt{2}-1$.

\begin{figure}
\vspace*{4ex}
\centering
\includegraphics[scale=0.4]{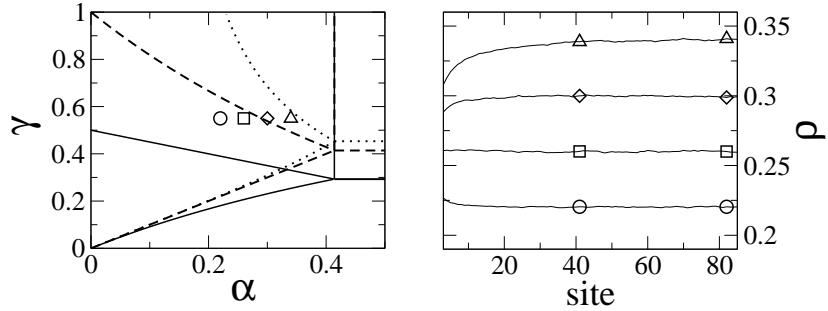}
\caption{\label{peak_decay}In the left-hand plot, the simple (solid
  line), refined (dashed line) and further refined (dotted line) phase
  boundaries are shown.  Simulations were carried out in the low
  density phase, across the predicted boundary between profiles with a
  decay from above and below $\alpha$ (parameter values indicated with
  symbols).  The resulting profiles, shown in the right-hand plot,
  display a transition that is in best agreement with the refined
  mean-field transition line.  }
\end{figure}

Further evidence in support of the refined mean-field theory can be
found by exploring the density profiles of the low density region,
where the three theories predict different transitions between
profiles which decay at the tip from below and above $\alpha$.  In
Figure~\ref{peak_decay}, we see that the transition appears to agree
with that predicted by the first refined theory, while being at odds
with both the further refined and the simple approximations.

\begin{figure}
\centering
\includegraphics[scale=0.4]{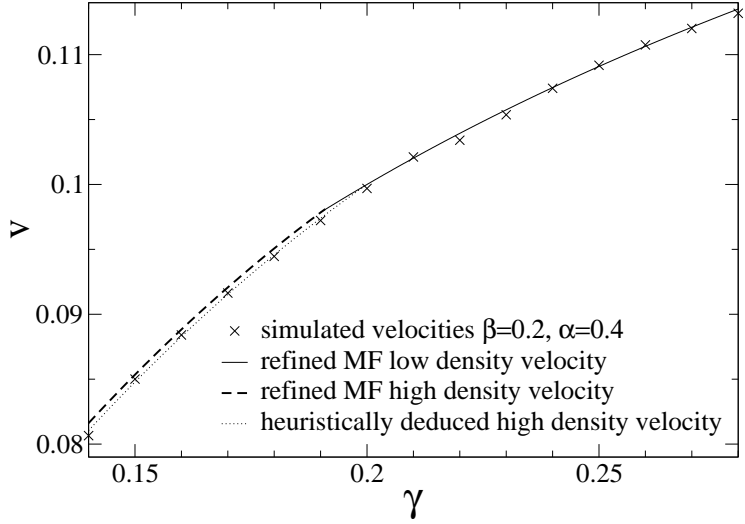}
\caption{\label{fig:withBeta} Theoretical and simulation results for
the tip velocity as a function of $\gamma$, given for fixed values of
$\alpha$ and $\beta>0$.  The data points show simulation results,
fitted to the heuristic and refined mean-field functions for the tip
velocity in the high and low density phases.  The low density velocity
is the same for heuristic and mean-field approaches and is given by
(\ref{vld}).  The high density velocity function is found in the
refined mean-field theory by solving equation (\ref{quadratic}) and
inequality $\rho_3<1-v-\alpha$ (explicit expression for the high
density velocity is not presented here). The heuristic high
density velocity is given by (\ref{v_heur}).  The
phase transition occurs at approximately $v_{ld}=v_{hd}$}
\end{figure}

For completeness, we consider now the case of non zero $\beta$.  In
Figure \ref{fig:withBeta} the simulated growth velocities are compared
to the refined mean-field theory velocity functions for the high and
low density phases for $\beta\neq0$.  The high density velocity
obtained from the heuristic argument is also shown and one may see
that this coincides rather closely with the refined mean-field
result. Due to the nature of Equations \ref{quadratic} and inequality
$\rho_3<1-v-\alpha$ which must be solved simultaneously, in this case
it is simpler to work with the velocity as a function of $\gamma$.
Therefore, simulations were carried out with fixed values $\alpha=0.4$
and $\beta=0.2$, in the range close to the phase transition
$\gamma=0.14..0.28$.  We see a good agreement of the simulation data
points with the high density and the low density velocity functions
predicted by the refined mean field theory.  Note however, that in the
high density phase the simulation results appear after all to be in
best agreement with the heuristic result for the velocity
(\ref{v_heur}).  The phase transition occurs as expected where the
high and low velocity functions meet.

\begin{figure}
\vspace*{4ex}
\centering
\includegraphics[scale=0.4]{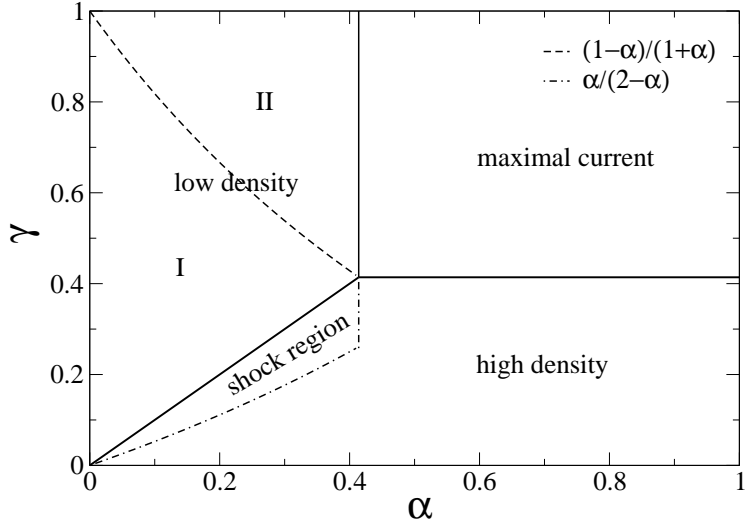}
\caption{\label{impPD}The phase diagram obtained from the refined
  mean-field approximation in the limit $\beta=0$.  Three phases are
  shown.  The low density phase is split into subregions I and II,
  indicating regions where the density profile decays from above and
  below the bulk density respectively.  The shock region is indicated
  within the high density phase.}
\end{figure}

Finally, we present the detailed 2-D phase diagram of the first
refined mean-field theory in the limit $\beta=0$, shown in
Figure~\ref{impPD}.  The diagram is split into the three distinct
phases: high, low and maximal current.  The low density phase is split
into two subregions I and II, where the tip decays to the bulk density
from above and below respectively.  The shock region, where the shock
recedes from both boundaries, is indicated within the high density
phase.  We believe this to be at least an accurate approximation to
the exact phase structure as we discuss in the next section.

\section{Conclusion}
\label{sec:conc}
In this work we have studied a generalisation of the open boundary
TASEP to a dynamically extending lattice and characterised a
generalised phase diagram in the $\alpha, \beta, \gamma$ parameter
space. The model provides an example of a system where the dynamics of
the microscopic constituents couple to and determine the dynamics of
the containing structure.  We conclude that steady states can exist in
this system, where the lattice is continuously extending from one end
while being supplied with particles from the other end.

The well-known phases of the open boundary TASEP are reflected in the
phase diagram of this model (Figure~\ref{impPD}).  The low density
phase has structure in the density profile near the tip and the bulk
density is controlled by the low input rate at the opposite
boundary. In the high density phase the bulk density is controlled by
the tip dynamics and there is structure in the density profile near
the opposite boundary.  In between we have a shock region where a high
density and a low density region are separated by a moving shock which
in fact recedes from both boundaries.  The shock region of the phase
diagram extends the known phases and adds to the variety
of shock formation phenomena which have been observed in the TASEP and
related models\cite{PFF,EJS,PRWKS,MB}.  Finally, the maximal current
phase is present but with bulk density $<1/2$.  However, in this model
the maximal current phase is not strictly a steady state as the
current and the density profile continue to evolve as the system
grows.

We have used several mean-field approximations and find that while the
simplest mean-field approach successfully predicts the qualitative
behaviour of the phases, the phase boundaries are not accurately
predicted.  However, a refined mean-field approximation which takes
into account the correlation between sites one and two predicts a
phase diagram which appears to coincide very closely with simulation
results and is very close to that deduced by the heuristic argument of
Section~\ref{sec:heur}. It would be of interest to know whether in fact
the predictions, summarised for the $\beta=0$ case in
Figure~\ref{impPD}, could be exact.  Interestingly, taking into
account a further correlation appeared to worsen the predicted phase
diagram. This sort of negative result has been encountered in
mean-field approximation schemes in the past \cite{bAK92}. For example
it was found in the ZGB surface reaction model that a pair mean-field
approximation successfully locates a discontinuous phase transition,
while a four-site approximation seems to do worse \cite{ron}.

The DEEP was initially formulated as a model derived from the TASEP of
a biophysical growth process, namely fungal hyphal growth \cite{SEPR}
and provides a simple framework which may find application in a
broader context.  The TASEP has been successfully developed to
describe various biophysical transport problems by incorporating
additional biological detail such as dynamic instabilities and
heterogeneity \cite{DL93,AKR07,EHK,KKCJ,Yariv}.  It could prove
fruitful to study the effects of such features in the present model.

\ack We would like to thank Wilson Poon, Nick Read and Graham Wright for their
encouragement and discussions on fungal biology, and Graeme Ackland
for helpful comments. KEPS thanks the EPSRC for funding.





\end{document}